\documentclass[12pt]{article}

\usepackage{jheppub}
\usepackage{jheppub}

\usepackage{amsmath,amssymb,array,mathrsfs,amsfonts,arydshln}
\usepackage{epsfig}

\newcommand\nn{\nonumber}
\newcommand{\be}{\begin{eqnarray}}
\newcommand{\ee}{\end{eqnarray}}
\newcommand{\m}{\mu}
\newcommand{\n}{\nu}
\newcommand{\al}{\alpha}
\def\l{\lambda}
\def\r{\rho}
\def\D{\Delta}
\newcommand{\x}{\chi}
\newcommand{\tr}{{\rm Tr}}
\def\VEV#1{\left\langle #1\right\rangle}
\def\str{{\rm Str}}
\def\sdet{{\rm Sdet}}
\newcommand{\cev}[1]{\reflectbox{\ensuremath{\vec{\reflectbox{\ensuremath{#1}}}}}}

\def\Dslash{\,\,{\raise.15ex\hbox{/}\mkern-12mu D}}
\usepackage{bbold}

\title{The gradient flow of the Dirac spectrum}
\author[a]{Alexander S. Christensen,}
\author[b]{K. Splittorff,}
\affiliation[a,b]{Discovery Center, The Niels Bohr Institute, University of 
Copenhagen, Blegdamsvej 17, 2100
Copenhagen, Denmark}
\author[c]{and J.J.M. Verbaarschot}
\affiliation[c]{Department of Physics and Astronomy, State University of New York at Stony Brook, NY 11794-3800, USA}
\abstract{
We construct chiral perturbation theory for the gradient 
flow of the microscopic Dirac eigenvalues and compute 
the density of and correlations between the microscopic 
eigenvalues at zero and non-zero flow time. The results show 
that the repulsion of the microscopic Dirac eigenvalues from 
the dynamical quark mass decreases with increasing gradient flow time.
Furthermore, the flow of the spectral resolvent is compared to the flow of the chiral
condensate obtained from a fermionic gradient flow.}

\begin{document}

\maketitle 
\newpage

%%\tableofcontents

\section{Introduction}

The phenomelogical success of Lattice QCD relies on a continuum extrapolation 
which is consistent with the renormalization properties of the strong interactions.
To reduce the ultraviolet quantum fluctuations and to improve the continuum 
extrapolation, it is not uncommon to smoothen the gauge field configurations on the scale 
of several lattice spacings. Recently the gradient flow \cite{luscher-1,luscher-2,luscher-weisz,luscher-chiral} 
has provided new methods to determine renormalized quantities on the lattice, see for example \cite{Suzuki,DelDPR,RS,Daniel}.
 The extension of the gradient flow to include the flow of the quark fields \cite{luscher-chiral},
raises a natural question: {\sl To which extend does  gradient flow 
preserve the dynamical properties induced by the fermion determinant?} As a first step towards 
a complete answer we  compute in this paper the effect of gradient flow on the microscopic 
eigenvalues of the Dirac operator. 
In a standard dynamical simulation the fermion determinant induces an eigenvalue repulsion 
from the quark mass  which increases for decreasing quark mass. This 
affects the microscopic eigenvalue density in a universal way \cite{DN,WGW,DOTV} which has been verified 
by  lattice simulations with dynamical quarks \cite{BerbenniBitsch:1998sy,Damgaard:2000qt,Fukaya}. 
The microscopic eigenvalue density at nonzero gradient flow time, computed here, gives the flow time 
scale for which the microscopic correlations induced by the fermion determinant persist.

Ideally, we would like to see that the large eigenvalues of the Dirac operator 
flow beyond the ultraviolet 
cut-off so that they do not contribute to the chiral condensate, while the small eigenvalues 
remain invariant under the flow such that their dynamical properties stay intact.  
From the calculation of the chiral condensate \cite{luscher-chiral,BG,shindler}
one might think that the flow only affects the low-lying eigenvalues by a trivial 
rescaling. In practice,
the situation will be more complicated: We will show that  gradient flow decorrelates the microscopic eigenvalues of the
Dirac operator from the eigenvalues at zero flow time. In particular, the microscopic
spectral density evolves from the universal  form for dynamical quarks, which includes the 
correlations induced by the fermion  determinant, to the quenched 
spectral density. This transition occurs for flow times of order $1/\sqrt{V}$, where a 
new low energy constant makes the dimensions match.

All results will be obtained from the chiral Lagrangian for the 
generating function of the Dirac spectrum in the $\epsilon$-domain of QCD. 
This generating function has dynamical 
quarks at zero flow time while the valence quarks are at non-zero flow time. The 
resulting flavor symmetry breaking in the generating function gives rise to 
 a non-trivial 
$t-$dependence of the chiral Lagrangian for  gradient flow. As we will explain 
in detail, this explicit $t$-dependence is distinct from that of the 
chiral Lagrangian for the fermonic flow discussed in \cite{BG}. 

 Both the chiral condensate as defined through the fermonic flow and the spectral resolvent, 
 evaluated at the dynamical quark mass, evolve from the chiral condensate at zero flow 
 time. The evolution of the two is however different. We explicitly compute and compare 
 the two at non-zero flow time.

The outline of this paper is as follows. In section \ref{sec:flowdef} we give a basic
review of gradient flow. Chiral perturbation theory including the chiral
Lagrangian in the $\epsilon$-domain is discussed in section \ref{sec:flowCPT}.  Explicit results of the flow of the
one-flavor and quenched Dirac spectrum are evaluated in section \ref{sec:rho}.
Concluding remarks are made in section \ref{sec:conc}, and additional technical
details are worked out in the Appendix.

%======================================================
\section{The flow equations}
\label{sec:flowdef}

A fully dynamical lattice simulation of the strong interactions provides an ensemble of gauge 
field configurations, $A_\mu$, that includes correlations induced by the fermion determinant. As a way to reduce the ultraviolet fluctuations of these configurations 
one may consider instead the flow time ($t$) dependent gauge fields $B_\m(t,x)$ introduced in \cite{luscher-1}. 
They evolve from the conventional gauge fields $A_\m(x)$ through initial conditions $A_\m(x) = B_\m(0,x)$ and satisfy the first order differential equation
\be
\partial_t B_\m = D_\n G_{\m\n} \, ,
\label{FlowEqnG}
\ee
where $G_{\m\n}$ is the field strength tensor
\be
G_{\m\n} = \partial_\m B_\n - \partial_\n B_\m + [B_\m , B_\n] \, ,
\ee
and $D_\mu = \partial_\m + [B_\m , \cdot \, ]$ is the covariant derivative. Note that the gauge field flow 
is determined by the gradient of the Yang-Mills action and that the flow time has dimension of $L^2$ \cite{luscher-1}.
One may also introduce the quark field $\x(t,x)$ 
and anti-quark field $\bar{\x}(t,x)$ which flow according 
to \cite{luscher-chiral}
\be
\partial_t \x = \D \x \, , \quad \partial_t \bar{\x} = \bar{\x} \cev{\D} \, ,
\label{FlowEqnQ}
\ee
where $\D = (\partial_\mu+B_\mu)(\partial_\mu+B_\mu)$ is the covariant Laplacian. The initial conditions $\psi(x) =\x(0,x)$ and $\bar{\psi}(x) =\bar{\x}(0,x)$ relate the flow time dependent quark fields to the dynamical quark fields in the simulation at zero flow time.

In the above differential equations the quark fields at non-zero 
flow time do not enter the gauge field flow. Initially, at $t=0$, the 
quarks and gluons are dynamical, but at non-zero $t$ the gauge field 
flow is independent of the flow of the fermion. The central aim of 
this paper is to compute the flow of the Dirac eigenvalues and to discuss 
how the dynamical properties of the eigenvalues and the chiral 
condensate changes with flow time. 
In this discussion the central observables are:
\bigskip 

\noindent
{\bf 1)} The chiral condensate as obtained from the fermionic flow
\be
\left\langle S^0_t\right\rangle \equiv \left\langle\bar{\chi}\chi\right\rangle = \frac{1}{Z_{t=0}(m)}\int {\cal D}A_\mu{\cal D}\bar\psi{\cal D}\psi \; \bar\chi\chi  \; e^{\bar\psi(D(A_\mu)+m)\psi-S_{\rm YM}(A_\mu)}, 
\ee
where $\bar\chi$ and $\chi$ are obtained from $\bar\psi$, $\psi$ and $A_\mu$ through (\ref{FlowEqnQ}) and (\ref{FlowEqnG}). $D(A_\mu)$ is the Dirac operator evaluated on the gauge field $A_\mu$, $S_{\rm YM}$ is the standard Yang-Mills action and 
\be
Z_{t=0}(m) = \int {\cal D}A_\mu{\cal D}\bar\psi{\cal D}\psi \; e^{\bar\psi(D(A_\mu)+m)\psi-S_{\rm YM}(A_\mu)}. 
\label{Zt0}
\ee
Note that a new source term at time $t$ must be introduced into the generating 
function in order to write $\left\langle\bar{\chi}\chi\right\rangle$ as a 
derivative
\be
\left\langle S^0_t\right\rangle \equiv \left\langle\bar{\chi}\chi\right\rangle = \frac{1}{Z_{t=0}(m)}\lim_{m_t\to0}\frac{d}{dm_t}\int {\cal D}A_\mu{\cal D}\bar\psi{\cal D}\psi \; e^{\bar\psi(D(A_\mu)+m)\psi+m_t\bar\chi\chi-S_{\rm YM}(A_\mu)} \, . 
\label{genfct-chibarchi}
\ee
\bigskip

\noindent
{\bf 2)} The second kind of observables we will discuss are 
resolvents for the Dirac spectrum. 
In this case there is no gradient flow of the fermion fields and only the 
flow of the gauge fields
is taken into account.
The resolvent for the one point 
function of the Dirac spectrum at flow time $t$ is given by   
\be\label{genfct-resolvent}
\left\langle {\cal S}^0_t\right\rangle &\equiv& \left\langle \tr \frac{1}{D(B_\mu)+m_v} \right \rangle \\
& \equiv & \frac{1}{Z_{t=0}(m)}\lim_{\tilde{m}_v\to m_v}\frac{d}{dm_v} 
  \int {\cal D}A_\mu \; {{\det}}^{N_f}(D(A_\mu) + m) \frac{{\det}(D(B_\mu) + m_v)}{{\det}(D(B_\mu) + \tilde{m}_v)}e^{-S_{\rm YM}(A_\mu)} \, , \nn
\ee
where $B_\mu$ is obtained from $A_\mu$ through (\ref{FlowEqnG}). We discuss similarities and differences between $\left\langle {\cal S}^0_t\right\rangle$ and $\left\langle S^0_t\right\rangle$ in section \ref{sec:S_vs_calS}.  
The eigenvalue density of $D(B_\mu)$ will be obtained from $\left\langle {\cal S}^0_t\right\rangle$ in section \ref{sec:rho1pf}.

\bigskip

The computation of the above observables will be performed within 
chiral perturbation theory.

%======================================================
\section{Chiral Perturbation Theory for  Gradient Flow}
\label{sec:flowCPT}

Chiral perturbation theory at non-zero flow time in the $p$-regime for 
 $\left\langle S^0_t\right\rangle$ was introduced in \cite{BG}. 
After a brief review of these results and the extension thereof 
to the $\epsilon$-regime we set up chiral perturbation theory for 
 $\left\langle {\cal S}^0_t\right\rangle$. 

\subsection{$p$-regime and the generating function 
for $\left\langle S^0_t\right\rangle$}
\label{sec:p-regSandP}

To order the terms in the chiral Lagrangian a counting scheme must be adopted. 
We will start out with the $p$-regime counting  
\be
\partial_\m \thicksim \frac{1}{L} \thicksim \epsilon \, , \quad \text{and} \quad m \thicksim m_t \thicksim \frac{1}{\sqrt{V}} \thicksim \epsilon^2 \, .
\label{p-counting}
\ee
Note that the source $m_t$ will be put to zero after differentiation.
For the standard $t=0$ partition function (\ref{Zt0}) the chiral Lagrangian 
to leading order is simply given by \cite{GLorig,GL}
\be
{\mathcal L}=\frac{f^2}{4} \tr(\partial_\mu U \partial_\mu U^\dag) + \frac{\Sigma}{2} \tr({\cal M} U + U^\dag {\cal M}) \, ,
\label{ChLagr}
\ee
where  $\mathcal{M} = \mathrm{Diag}(m_1,\ldots,m_{N_f})$ is the quark mass 
matrix, the low energy constant $\Sigma$ is the chiral condensate and $f$ is 
the pion decay constant. This celebrated chiral Lagrangian describes the non-perturbative 
low energy phase where the strong interactions spontaneously break chiral symmetry.

The unitary field $U$ is defined as
\be
U(x) = \exp \left( \frac{2i \pi^a(x) T^a}{f} \right) \, ,
\ee
where $\pi^a$ are Goldstone bosons and $T^a$ are the generators of the $SU(N_f)$ flavor group.

The extension of the chiral Lagrangian needed to compute
$\langle S^0_t\rangle=\langle\bar\chi\chi\rangle$ at non-zero flow 
time was written down in \cite{BG}: As observed above 
Eq.~(\ref{genfct-chibarchi}) the generating function for 
$\langle S^0_t\rangle$ includes an explicit source term for this observable.
Since the chiral transformation properties of 
$\bar\chi$ and $\chi$ are independent of $t$, the spurion properties of the 
new quark mass source $m_t$ will be identical to that of the standard quark mass $m$. Hence, 
the $m_t$ dependent terms in the chiral Lagrangian must be of exactly 
the same form as the standard mass terms, but the magnitude of the low 
energy constants appearing in combination with $m_t$ can change with $t$. 
As the counting for $m$ and $m_t$ is identical, cf.~(\ref{p-counting}), 
the new term at leading order must have the same form as the original one 
\cite{BG}
\be
\mathcal{L}'=   \frac{m_t\Sigma_t}{2} \tr(U + U^\dag) \, ,
\ee
where we have introduced the low energy constant at flow time $t$, 
$\Sigma_t$. 
The sum ${\mathcal L}+\mathcal{L}'$ makes up 
the lowest order chiral Lagrangian in the generating function for 
$\langle S^0_t \rangle$ at leading order in the $p$-regime \cite{BG}.

Within the $p$-counting the Compton wavelength of the pions is comparable 
to the dimension of the box $1/M_\pi\thicksim L$.
Perturbatively, the flow introduces a smearing width of $\sqrt{8t}$ 
\cite{luscher-1}, 
and it is natural to consider this to be much smaller 
than $L$ and the Compton wavelength of the pions \cite{luscher-chiral,BG}.

\subsection{$\epsilon$-regime generating function for $\left\langle S^0_t\right\rangle$}
\label{sec:ep-regSandP}

In this section  we set up the $\epsilon$-regime of the generating function
for
$\left\langle S^0_t\right\rangle$ at non-zero flow time.
The natural extension of the counting in the $\epsilon$-regime is one where 
the standard mass $m$ and the source mass, $m_t$, for $\bar\chi\chi$ are of the same order 
\be
\partial_\m \thicksim \frac{1}{L} \thicksim \epsilon \, , \quad m \thicksim m_t \thicksim \frac{1}{V} \thicksim \epsilon^4 \, .
\label{epsilon-counting}
\ee
Also in this case the source $m_t$ will be put to zero after differentiation.
As in the $p$-regime the flow does not break the chiral symmetries and 
the form of the possible mass terms in the chiral Lagrangian are identical.

Note that the new low energy parameter $\Sigma_t$ evolves with $t$ from 
the value $\Sigma$ at $t=0$ in a way that is not determined by chiral perturbation 
theory.  In order to keep this constant of order 
$\epsilon^0$ we need to include the new scale set by the smearing width 
$\sqrt{8t}$ \cite{luscher-chiral} into the 
$\epsilon$-counting scheme. As in \cite{BG} we will consider the 
case where $\sqrt{8t}$ is smaller than the pion Compton wavelength 
\be
\sqrt{8t} \ll \frac{1}{M_{\pi}} \, .
\label{constraint}
\ee 
In the $\epsilon$-regime, however, this is condition is much less 
constraining since 
\be
1/M_{\pi} \thicksim 1/\epsilon^2 \thicksim L^2 \ .
\ee
For example, it would allow for a smearing width that it is comparable 
${\cal O}(1/\epsilon)$ or even larger than the extent of the box. 
This is, however, undesirable: If we expand the difference between 
the chiral condensate at zero and non-zero flow time for small $t$ 
we get \cite{BG}
\be
\Sigma_t - \Sigma = (t \Lambda_{\rm QCD}^2) \sigma_1 + 
(t \Lambda_{\rm QCD}^2)^2 \sigma_2 + ... \, ,
\label{taylor}
\ee 
where $\sigma_1,\sigma_2,...$ are the Taylor-coefficients. With  
$\sqrt{8t} \thicksim 1/\epsilon$ (which is allowed by the constraint (\ref{constraint})) this would 
imply that $\Sigma$ and/or $\Sigma_t$ would have to scale at least 
as fast as $1/\epsilon^2$. To keep $\Sigma_t$ of order one requires
a counting scheme with
\be
\sqrt{8t} \thicksim \epsilon^0 \quad \quad  \quad \quad (t \ {\rm counting \ for} \ \left\langle S^0_t\right\rangle).
\label{eps0}
\ee
Having settled the counting we can now write down the form of the 
effective theory for the generating function Eq.~(\ref{genfct-chibarchi}). 

In the $\epsilon$-regime the zero modes of the Goldstone field dominate the 
partition function and the kinetic term in \eqref{ChLagr} factorizes from the partition function
\cite{GL}. At leading order in the $\epsilon$-counting the effective 
generating function of $\left\langle S^0_t\right\rangle$ is hence is given by the 
group integral
\be
\mathcal{Z}_{N_f} & = & \int\limits_{U(N_f)}\hspace{-3mm}\mathcal{D}U \ {{\det}}^\nu(U) \exp \left(\frac{\Sigma V}{2} \tr \left[ \mathcal{M}U + U^{-1}\mathcal{M} \right]+\frac{m_t\Sigma_tV}{2}\tr \left[U + U^{-1}\right] \right), \nn\\
\label{Zgen-chibarchi}
\ee
where $V$ is the four volume. The integration is over $U(N_f)$ since we consider the sector with fixed 
topological index $\nu$ \cite{LS}. Note that the $t$-dependence enters through the undetermined flow time 
dependent low energy parameter, $\Sigma_t$, and that $\sqrt{8t}\thicksim \epsilon^0$. (For 
$t\thicksim\epsilon$ one may replace $\Sigma_t$ by $\Sigma$ in (\ref{Zgen-chibarchi}), since the difference 
between the two will be of order $\epsilon$ acceding to (\ref{taylor}).)

\bigskip

Let us now turn to the effective generating function for the spectral 
resolvent of Eq.~(\ref{genfct-resolvent}). The presence of determinants 
at different flow times presents a new situation not faced 
in \cite{BG}. 

\subsection{$\epsilon$-regime generating function for $\left\langle {\cal S}^0_t\right\rangle$}
\label{sec:genfctcalSandP}

In this section we construct the low energy generating function for the spectral 
resolvent $\left\langle {\cal S}^0_t\right\rangle$. For simplicity consider 
one dynamical flavor where the generating function for 
$\left\langle {\cal S}^0_t\right\rangle$ is given by
\be\label{Nf1} 
&& \int {\cal D}A_\mu\, {\det}(D(A_\mu) + m)\frac{{\det}(D(B_\mu) + m_v)}{{\det}(D(B_\mu) + \tilde{m}_v)} \, e^{-S_{\rm YM}(A_\mu)} \\
&=& \int {\cal D}\bar\psi{\cal D}\psi{\cal D}\bar\eta_f{\cal D}\eta_f{\cal D}\bar\eta_b{\cal D}\eta_b {\cal D}A_\mu \, e^{-S_{\rm YM}(A_\mu)}\; \nn\\
&&\hspace{1cm} \times \exp\left[(\bar\psi \ \bar\eta_f \ \bar\eta_b)\left(\begin{array}{ccc} D(A_\mu)+m & 0 & 0 \\ 0& D(B_\mu)+m_v & 0\\ 0 & 0 & D(B_\mu)+\tilde{m}_v \end{array}\right)\left(\begin{array}{c} \psi \\ \eta_f \\ \eta_b \end{array}\right)\right].
\nn
\ee
We stress that the gauge fields $B_\mu$ are obtained from (\ref{FlowEqnG}) using the dynamical gauge field $A_\mu$ as initial 
condition while the $\eta$ fields are valence quarks. Note also that $\eta_b$ and $\bar\eta_b$ obey bosonic statistics.

Since the Dirac operators are not defined for the same gauge field 
configurations the flavor symmetries are violated 
even at vanishing quark masses. 
The difference between the Dirac operators leads to a new term 
of the form 
\be
\gamma_\mu (B_\mu-A_\mu) {\rm Diag}(0,1,1). 
\ee
Note that this is not an external source term, $B_\mu$ depends on $A_\mu$ 
and will be integrated out along with it.

With infinitesimal $t$ we have that  cf.~(\ref{FlowEqnG})
\be
B_\mu-A_\mu = D_\nu F_{\mu\nu} t \, ,
\ee
where $F_{\mu\nu}$ is the field strength at $t=0$ and $D_\mu$ likewise is evaluated at $t=0$. 
Hence, the new term in the generating function for small $t$ is
\be
\bar\Psi \;  t \; {\rm Diag}(0,1,1) \gamma_\mu D_\nu F_{\mu\nu} \; \Psi,
\label{new-term}
\ee 
where
\be
\Psi = \left(\begin{array}{c} \psi \\ \eta_f \\ \eta_b \end{array}\right).
\ee

The new term, (\ref{new-term}), breaks the flavor symmetries as a vector 
source (this linearized breaking is sufficient as we will work to leading order in $t$).
 Hence, in the chiral Lagrangian for the  
generating function we must include all 
possible terms which break the symmetries in this manner. To identify 
these terms we promote $t$ to a spurion field which 
transforms such that (\ref{new-term}) is invariant. In the chiral Lagrangian 
theory we then include 
all the possible invariant terms including the spurion $t$. The possible terms with at most two $t$'s are 
\be
\str\left[t U^{-1}t U\right], \quad \str\left[t t\right] \quad {\rm and} \quad 
i\str\left[t_\mu U^{-1}\partial_\mu U - t_\mu (\partial_\mu U^{-1})U\right],
\label{possible-terms}
\ee
with $t_\mu=t$.

To obtain the leading terms in the chiral Lagrangian  we need to 
define a counting scheme. We will adopt $t$ into the $\epsilon$-counting 
such that 
\be
\partial_\m \thicksim \frac{1}{L} \thicksim \epsilon \, , \quad \text{and} \quad m \thicksim m_v \thicksim \frac{1}{V} \thicksim \epsilon^4 \quad \text{and} \quad t \thicksim \frac{1}{L^2} \thicksim \epsilon^2 \nn\\ 
  \quad \quad  \quad \quad \quad \quad \quad \quad (t \ {\rm counting \ for} \ \left\langle {\cal S}^0_t\right\rangle)
  \, .
\label{epsilon-counting-incl-t}
\ee
Note that the counting for $t$ is different from that in the generating function for $\left\langle S^0_t\right\rangle$.
Then to leading order we have the low energy generating function\footnote{The rightmost term in (\ref{possible-terms}) 
only produces an irrelevant boundary term, for a related discussion see the introduction of \cite{DHSST}.}
\be\label{ZNf_1pf}
 && {\cal Z}_{N_f+1|1}(m,m_v|\tilde{m}_v) \\
 & = & \int\limits_{Gl(N_f+1|1)}\hspace{-3mm}  \mathcal{D}U \ {\sdet}^\nu U \exp \left(\frac{1}{2} \str \left[\mathcal{M}U + U^{-1}\mathcal{M}\right] +t^2 F_{\rm flow}^2 V \str\left[{\cal T}U^{-1}{\cal T}U\right] \right) \, , \nn
\ee
where $\mathcal{M} = \mathrm{Diag}(\overbrace{m\Sigma V,\ldots,m\Sigma V}^{N_f},m_v\Sigma V,\tilde{m}_v \Sigma V)$ and ${\cal T}={\rm Diag}(\overbrace{0,\ldots,0}^{N_f},1,1)$.

The condensate $\Sigma$ that multiplies $m_v$ differs from the condensate
at $t=0$ by terms of $O(t)$ which do not contribute to leading order 
in the $\epsilon $ counting. The  new low energy constant
$F_{\rm flow}$ depends on the details of  
the flow. Other flow equations for $B_\mu$ which break the symmetries 
of the generating functional in the same manner as the flow considered 
here, will correspond to different values of $F_{\rm flow}$. Note that $F_{\rm flow}$
has dimension 4 such that $t^2 F_{\rm flow}^2 V$ is dimensionless. 

\subsection{The similarities and differences between $\langle S^0_t\rangle$ and $\langle {\cal S}^0_t\rangle$}
\label{sec:S_vs_calS}

Since $\langle S^0_t\rangle$ and $\langle {\cal S}^0_t\rangle$ (evaluated at the physical quark 
mass) both evolve from $\langle\bar\psi\psi\rangle$ at zero flow time one may potentially use either to 
define the chiral condensate at non-zero flow time. While $\langle S^0_t\rangle$ is defined through the 
flow of the fermions $\langle {\cal S}^0_t\rangle$ depends only on the gradient flow  of the gauge fields.
The two are therefore not necessarily equal at nonzero flow time. 
To compare the two quantitatively we here 
discuss their exact form in the $\epsilon$-regime. 
\bigskip

\begin{figure}[t!]
\centerline{\includegraphics[width=0.75\textwidth]{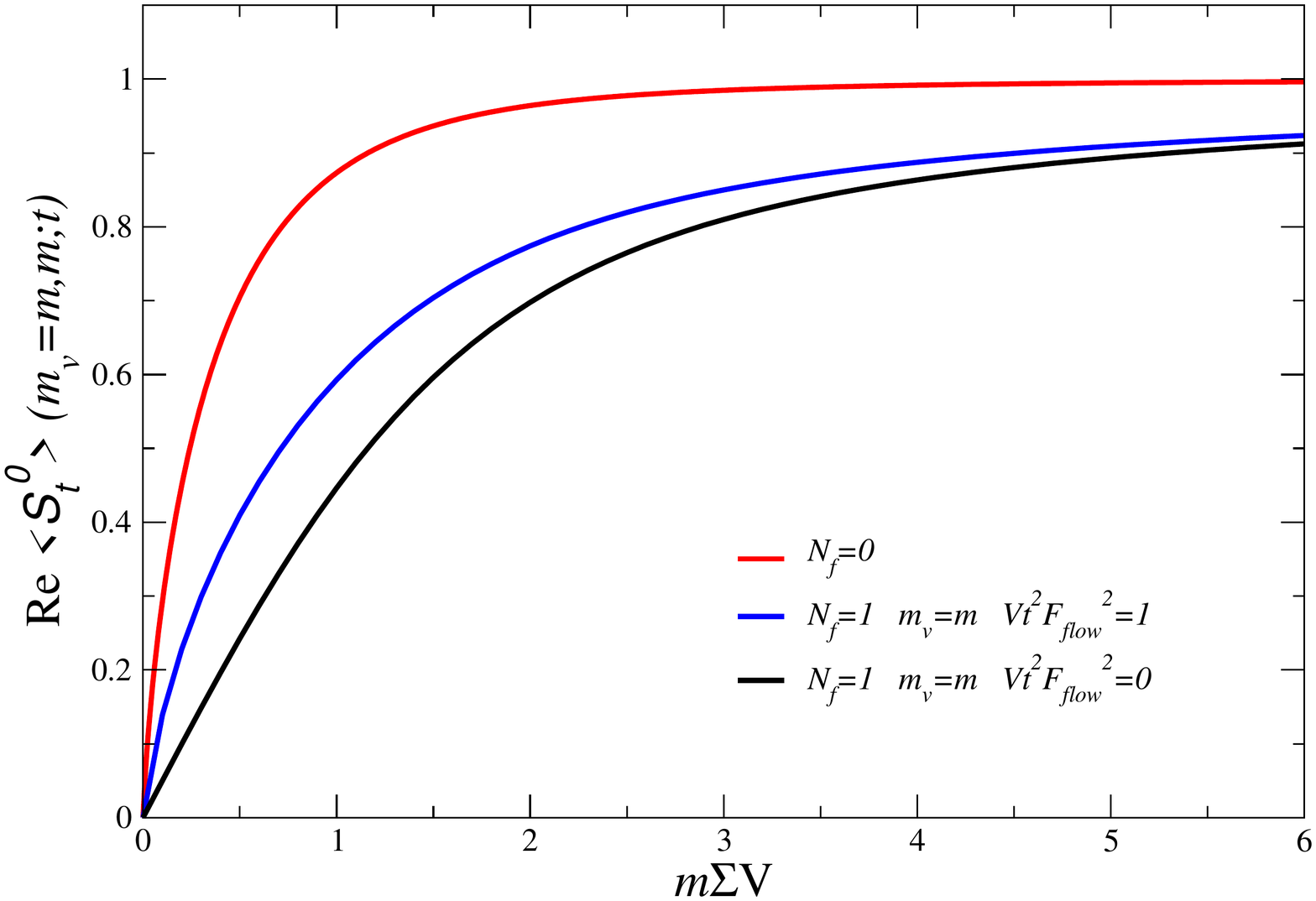}}
\caption{\label{fig:SigmaNf1} The real part of the spectral resolvent 
at flow time $F_{\rm flow}^2t^2V=1$ (middle curve) evaluated at the 
valence mass equal to the dynamical mass, $m_v=m$, and plotted as a 
function of $m\Sigma V$. At $t=0$ the result (lower curve) matched the 
fully dynamical microscopic eigenvalue density while the quenched 
condensate (upper curve) is approached for large flow time 
$F_{\rm flow}^2t^2V\gg1$.} 
\end{figure}

In the $\epsilon$-regime the condensate from the fermonic flow, 
$\langle S^0_t\rangle$, follows from (\ref{Zgen-chibarchi}) by evaluating 
the derivative wrt.~$m_t$ of $\log Z$ at $m_t=0$. From 
(\ref{Zgen-chibarchi}) it is clear this expression takes the same form as the 
standard chiral condensate at $t=0$, only the overall scale is now set 
by the flow time dependent low energy parameter $\Sigma_t$. For $N_f=1$ 
the result is   
\be
\langle S^0_t\rangle = \Sigma_t\frac{1}{2}\frac{I_{\nu+1}(m\Sigma V)+I_{\nu-1}(m\Sigma V)}{I_\nu(m\Sigma V)} \ .
\label{cond-eps}
\ee
Note that the $t$ dependence of $\Sigma_t$ is not determined by chiral 
perturbation theory.

To determine $\VEV{{\cal S}_t^0}$ for $N_f=1$ in the $\epsilon$-regime we make use of an explicit 
parametrization of $Gl(2|1)$ in the generating function (\ref{ZNf_1pf}). The parametrization and details are given in Appendix \ref{app:parametrization}.
A plot of $\VEV{{\cal S}_t^0}$ evaluated at the 
valence quark mass, $m_v$, equal to the physical quark mass, $m$, as a function
of $mV\Sigma$ 
is shown in Fig. \ref{fig:SigmaNf1}. 
 Initially, at zero flow time, $\VEV{{\cal S}_t^0}$ is the dynamical condensate 
which includes the full effect of the fermion determinant. 
With increasing flow time 
$\VEV{{\cal S}_t^0}_{m_v=m}$ approaches the quenched form which it reaches
 in the limit $t^2F_{\rm flow}^2V\gg1$.

%======================================================
\section{Dirac spectra}
\label{sec:rho}

Let us now turn to the flow of the Dirac spectra. At zero flow time
the eigenvalues of the Dirac operator are determined by the eigenvalue
equation \be D(A_\mu) \psi^A_n = i \lambda^A_n \psi^A_n \, .  \ee As
the gauge field $B_\mu$ evolves with the flow according to
(\ref{FlowEqnG}) from its initial value $A_\mu$, the eigenvalues of
the Dirac operator evaluated on $B_\mu$ are given by 
\be 
D(B_\mu)\psi^B_n = i \lambda^B_n \psi^B_n \, .  
\ee 
Here we will compute the
microscopic eigenvalue density at time $t$ as well as the two point
function between the spectra at time $t=0$ and at time $t$.

\subsection{The spectral one point function in the $\epsilon$-regime}
\label{sec:rho1pf}

\begin{figure}[t!]
\centerline{\includegraphics[width=0.75\textwidth]{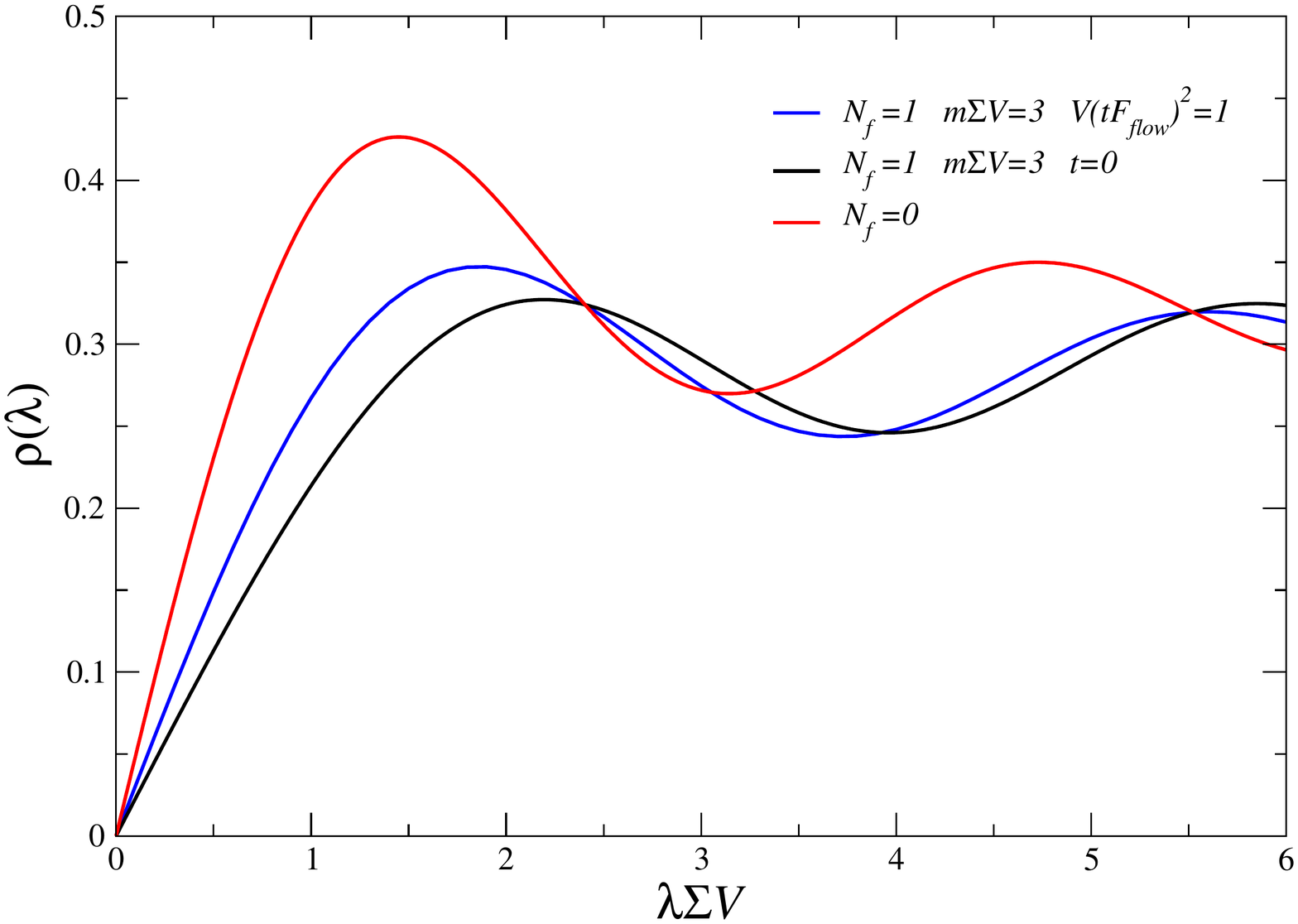}}
\caption{\label{fig:rhoNf1} The microscopic eigenvalue density for zero 
topological charge at non-zero 
flow time $t$. As the flow proceeds the microscopic eigenvalue density smoothly 
moves from the fully dynamical result (at $t=0$) to the quenched form at 
$F_{\rm flow}^2t^2V\gg1$. In the plot the dynamical quark mass is $m\Sigma V= 3$ 
and the  flow time is chosen such that $F_{\rm flow}^2t^2V=1$.} 
\end{figure}

Since $\langle{\cal S}_t^0\rangle$ is  the partially quenched 
condensate, the eigenvalue density of $D(B_\mu)$ is simply the discontinuity 
of $\langle{\cal S}_t^0\rangle$ across the imaginary axis
\be
\rho_t(\lambda^B,m) & = & \lim_{\epsilon\to0} \frac{1}{\pi}\left(\left\langle \tr \frac{1}{D(B_\mu)+i\lambda^B+\epsilon} \right \rangle - \left\langle \tr \frac{1}{D(B_\mu)+i\lambda^B-\epsilon} \right \rangle \right) \\
 & = & \lim_{\epsilon\to0} \left(\left \langle {\cal S}_t^0 \right \rangle_{m_v =i\lambda^B+\epsilon} - \left\langle {\cal S}_t^0 \right \rangle_{m_v =i\lambda^B-\epsilon}\right) \nn. 
\ee

\noindent
Let us first consider the quenched case. In this case the generating function, (\ref{Nf1}),
for $\left \langle {\cal S}_t^0 \right \rangle$ only involves the valence 
quarks. Both of these appear together with $D(B_\mu)$ and the evolution 
of the gauge field therefore does not break the flavor symmetries of the 
generating 
function. The $t^2$ term in the effective generating function is therefore 
absent. This  also follows directly from (\ref{ZNf_1pf}) by observing 
that the ${\cal T}$ matrix for $N_f=0$ is proportional to unity, resulting in 
\be
\str[U{\cal T}U^{-1}{\cal T}] \propto \str[UU^{-1}] =\str[{\mathbb 1}]= 0 \quad \quad \quad (N_f=0).
\ee
We conclude that the quenched microscopic eigenvalue density is independent 
of the flow time, and in the sector of topological charge $\nu$ it is given 
by \cite{VZ}
\be\label{rhoQ}
\rho_{N_f=0}(\lambda^B\Sigma V) = \frac{1}{2}\lambda^B\Sigma V\left[J_\nu^2(\lambda^B\Sigma V)-J_{\nu-1}(\lambda^B\Sigma V)J_{\nu+1}(\lambda^B \Sigma V)\right]+\nu\delta(\lambda^B \Sigma V).\nn \\
\ee

\noindent
The unquenched eigenvalue density of $D(B_\mu)$ depends on time as 
the ${\cal T}$ matrix is non-trivial. To 
obtain the unquenched eigenvalue density we only need
to put $m_v=i\lambda^B$ and take the real  part   of 
$\langle{\cal S}_t^0\rangle$ calculated in Appendix \ref{app:parametrization} 
for $N_f=1$.
The resulting eigenvalue density  
is plotted in Fig.~\ref{fig:rhoNf1} for 
$VF_{\rm flow}^2t^2=1$ and $m\Sigma V=3$. Also shown is the fully 
dynamical $t=0$ result as well as the quenched spectral density. 
At small $\lambda\Sigma V$ the fully dynamical eigenvalue density falls below the quenched 
curve due to the repulsion of the smallest eigenvalues from the dynamical quarks mass. 
The curve for $VF_{\rm flow}^2t^2=1$ shows how this repulsion is reduced by the flow such that  
at $VF_{\rm flow}^2t^2\gg 1$ the microscopic eigenvalue density takes the quenched form 
(recall that the new low energy constant $F_{\rm flow}$ has dimension 4 such that $VF_{\rm flow}^2t^2$ 
is dimensionless).
The flow time scale for the loss of these dynamical correlations is already much
earlier at  $t\thicksim m_\pi f/ F_{\rm flow}$.
 In 
this manner $F_{\rm flow}$ determines how long the microscopic eigenvalue 
spectrum maintains its dynamical properties.

Notice also that the number of eigenvalues equal to zero remains 
unchanged as the time flows. In other words, the index of $D(A_\mu)$ is 
equal to that of $D(B_\mu)$. 

%=================================================================

\subsection{The spectral two-point function in the $\epsilon$-regime}

In order to better understand the flow of the eigenvalues of the Dirac 
operator  we will here extend the standard spectral two point 
function $\r(\l_1,\l_2)$ (where both eigenvalues are part of 
the spectrum of the Dirac operator evaluated for $A_\mu$ 
at time $t=0$) to non-zero flow time. We will consider the case where  
$\lambda_1=\lambda^A$ is an eigenvalue of $D(A_\mu)$ at $t=0$ 
and $\lambda_2=\lambda^B$ is an eigenvalue of $D(B_\mu)$ at time $t$.
This spectral correlation function is denoted by $\r(\l^A,\l^B;t)$.
\bigskip

The correlation function between the eigenvalues at zero and non-zero $t$ reads 
\begin{align}
\begin{split}
\rho(\lambda^A,\lambda^B;t) \equiv&\, \VEV{\sum_n \delta(\lambda^A - \lambda^A_n) \sum_m  \delta(\lambda^B - \lambda^B_m)}\\
& - \VEV{\sum_n  \delta(\lambda^A - \lambda^A_n)} \VEV{\sum_m \delta(\lambda^B - \lambda^B_m)} \, .
\end{split}
\end{align}
It is related to the susceptibility through
\begin{equation}
\rho(\lambda^A,\lambda^B;t) = \frac{1}{4\pi^2} {\rm disc} \, \chi(m_A,m_B;t) \Big\vert_{\substack{m_A = i\lambda^A \\ m_B = i\lambda^B}} \, ,
\label{DiscChi}
\end{equation}
where
\begin{equation}
\chi(m_A,m_B;t) = \VEV{\sum_n \frac{1}{i\lambda_n^A + m_A} \sum_m \frac{1}{i\lambda_m^B + m_B}} - \VEV{\sum_n \frac{1}{i\lambda_n^A + m_A}} \VEV{\sum_m \frac{1}{i\lambda_m^B + m_B}} \, .
\end{equation}
A similar spectral two point function was used in 
\cite{DHSS,DHSST,Damgaard:2006pu} to extract the 
value of the pion decay constant from lattice QCD simulations.

Before performing an exact calculation of the two-point function, let 
us first give a qualitative discussion of its properties.
The two point correlation function satisfies  the identities
\be
\int \rho (\lambda^A,\lambda^B,t)\; d\lambda^A  &=& 0 ,\nn \\ 
\int \rho (\lambda^A,\lambda^B,t)\; d\lambda^B  &=& 0, 
\label{sumrule}
\ee
which follow immediately from the normalization
\be
\int\sum_n \delta (\lambda -\lambda^A_n)\; d\lambda = N
\ee
with $N$ equal to the total number of eigenvalues and the corresponding normalization for $\lambda^B_n$.
Let us first discuss the case of $t=0$. Then the first term in (4.6)
can be decomposed as
\be
\delta(\lambda^A -\lambda^B) \rho(\lambda^A) + \sum _{n\ne m}
\delta (\lambda^A - \lambda^A_n)\delta (\lambda^B - \lambda^B_m).
\ee
 Correlations due to the first term are known as self-correlations, whereas
the second term represents the genuine two-point correlations. Because 
of the sum rule (\ref{sumrule}) the integral over the genuine two-point
correlations is negative with the total negative area equal to the spectral
density  at $\lambda^A$. Since spectral correlations decrease for increasing
distance, we expect that the genuine two-point correlation function starts
at  a negative number and  then asymptotes to zero for increasing 
distance.

At small nonzero flow time, we expect that eigenvalues $\lambda^B_k$ 
fluctuate in a Gaussian way about the eigenvalues $\lambda^A_k$, i.e.
\be
\lambda^B_k = \lambda^A_k +\delta^B_k  
\ee
with the distribution of $\delta^B_k$ given by
\be
P(\delta^B_k) = \frac 1{\sqrt{2\pi\sigma}} e^{-(\delta^B_k)^2/2\sigma^2} .
\ee
For the self-correlations we now find
\be
&&\int d\delta^B_k P(\delta^B_k) \left \langle 
\sum_k \delta (\lambda^A - \lambda^A_k)\delta (\lambda^B - \lambda^A_k -\delta_k^B)\right \rangle \nn\\
&&= \frac{\rho(\lambda_A)}{\sqrt{2\pi\sigma}} e^{(\lambda^A-\lambda^B)^2/2\sigma^2}.
\ee 
So for the two-point correlation function at small nonzero flow-time, we
expect to find a Gaussian peak centered at $\lambda_A = \lambda_B$ and because of the
sum rule (\ref{sumrule}) a negative correlation gap away from this peak. 
This is exactly the behavior we will find in the explicit calculation in
the next section (see Fig. \ref{fig:two-pfQ}).

\subsubsection{Quenched two-point function}

In order to compute the spectral two point correlation function we will 
use a version of the replica method that was introduced in \cite{SVtoda}.
In this approach the replica limit is obtained from a recursion relation
for the generating function.  
We will consider the two-point function in the quenced case $N_f=0$. 

In replica approach, see 
eg.~\cite{rep-sk,replica,SVtoda,Kanzieper}, 
the susceptibility reads
\begin{equation}
\chi(m_A,m_B;t) \equiv \lim_{n \to 0}  \frac{1}{n^2} \partial_{m_A} \partial_{m_B}\log Z_n(m_A,m_B;t) \, ,
\label{ChiRep}
\end{equation}
where the replicated generating function is given by
\be
Z_n(m_A,m_B;t) =  \int{\cal D} A_\mu \, {\det}^n(D(A_\mu)+m_A)\, {\det}^n(D(B_\mu)+m_B)\, e^{-S_{\rm YM}(A_\mu)} \, .
\ee 
In the low energy effective description (cf.~section \ref{sec:genfctcalSandP}) 
we have
\begin{equation}
Z_n(m_A,m_B;t) = \int_{U(2n)} {\cal D}U \, {\det}^\nu(U) e^{ \frac{1}{2}\Sigma V \tr({\cal M}^\dag U +{\cal M} U^\dag) + VF_{\rm flow}^2 t^2 \tr({\cal T} U^\dag {\cal T} U)} \, ,
\label{ZnRep}
\end{equation}
with ${\cal M} = {\rm Diag}(m_A,\ldots,m_A,m_B,\ldots,m_B)$ and ${\cal T} = {\rm Diag}(0,\ldots,0,1,\ldots,1)$. 

In order to compute the low energy replicated generating function we first 
note that the $U(2n)$ integral changes only by a trivial factor if we shift 
${\cal T}$ by the unit matrix since 
\be
\tr(({\cal T}- \frac{1}{2}\mathbb{1})U^\dag ({\cal T}-{\frac{1}{2}}\mathbb{1}) U) 
= \tr({\cal T}U^\dag {\cal T}U)-\tr({\cal T})+\frac{1}{4}\tr\mathbb{1} = \tr({\cal T}U^\dag {\cal T}U)-\frac{1}{2}n\, . \nn \\
\ee 
Now using that $({\cal T}- \frac{1}{2}\mathbb{1})=-\frac{1}{2}B$, where 
$B\equiv{\rm Diag}(\mathbb{1},-\mathbb{1})$, the replicated partition 
function can be written
\begin{equation}
Z_n(m_A,m_B;t) = e^{\frac{1}{2}nVF_{\rm flow}^2t^2}\int_{U(2n)} {\cal D}U \, {\det}^\nu(U) e^{\frac{1}{2}\Sigma V \tr(M^\dag U + M U^\dag) + \frac{1}{4}VF_{\rm flow}^2 t^2 \tr(B U^\dag B U) } \, .
\label{ZnRep}
\end{equation}
The integral over $U(2n)$ above is of the same form as the one considered 
in \cite{SVfact,DHSS}, where it was found that  
the replicated partition function with $2n$ fermions can be expressed 
in terms of the  $n=1$ partition function
\begin{equation}
(m_Am_B)^{n(n-1)} Z_n(m_A,m_B;t) = D_n {\det} \left[ (m_A \partial_{m_A})^k(m_B \partial_{m_B})^l Z_1(m_A,m_B;t) \right] \, ,
\end{equation}
where $D_n$ is a constant to ensure normalization and $k,l = 0,1,...,n-1$. Since the above partition function 
is a $\tau$-function the $Z_n$ satisfy the Toda lattice equation \cite{DHSS}
\begin{align}
\begin{split}
\frac{1}{4n^2V^4\Sigma^4}m_A \partial_{m_A}m_B \partial_{m_B} \log Z_n&(m_A,m_B;t) \\
=&\, (m_A m_B)^2 \frac{ Z_{n+1}(m_A,m_B;t)  Z_{n-1}(m_A,m_B;t)}{\left[  Z_n(m_A,m_B;t) \right]^2} \, .
\end{split}
\end{align}
In the replica limit $n\to0$ the left hand side yields the susceptibility 
\eqref{ChiRep} apart from a constant. (The use of the integrable hierarchy 
for the replica limit was introduced in \cite{Kanzieper,SVtoda,SVfact}.) 
This implies that
\begin{equation}
\chi(m_A,m_B;t) = 4V^4\Sigma^4 m_A m_B  Z_1(m_A,m_B;t)  Z_{-1}(m_A,m_B;t) \, ,
\end{equation}
where 
\be
Z_1(m_A,m_B;t)  &=& e^{-2V F_{\rm flow}^2t^2} \int_0^1 {\rm d}\alpha \, \alpha e^{2V F_{\rm flow}^2t^2 \alpha^2} I_\nu(\alpha m_A \Sigma V) I_\nu(\alpha m_B \Sigma V) \, , \\
Z_{-1}(m_A,m_B;t) & =& e^{2VF_{\rm flow}^2 t^2} \int_1^\infty {\rm d}\alpha \, \alpha e^{-2VF_{\rm flow}^2 t^2 \alpha^2} K_\nu(\alpha m_A \Sigma V) K_\nu(\alpha m_B \Sigma V) \, . \nn
\ee
Here, $Z_{-1}$ is the partition function of two quarks with bosonic statistics. 
These expressions are analogous to those at non-zero imaginary isospin chemical 
potential, see \cite{DHSS} for details. 

The explicit expression for the susceptibility is thus given by
\begin{align}
\begin{split}
\chi(m_A,m_B;t) =&\, 4V^4\Sigma^4 m_A m_B \int_0^1 {\rm d}\alpha \, \alpha e^{2V F_{\rm flow}^2t^2 \alpha^2} I_\nu(\alpha m_A \Sigma V) I_\nu(\alpha m_B \Sigma V)  \\
&\times \int_1^\infty {\rm d}\alpha \, \alpha e^{-2VF_{\rm flow}^2 t^2 \alpha^2} K_\nu(\alpha m_A \Sigma V) K_\nu(\alpha m_B \Sigma V) \, .
\end{split}
\end{align}
The discontinuity thereof across the imaginary axis is the two point 
spectral correlation function cf.~\eqref{DiscChi},
\be
\label{twop}
\rho(\hat{\lambda}^A,\hat{\lambda}^B;t) &=& \hat{\lambda}^A \hat{\lambda}^B \int_0^1 {\rm d}\alpha \, \alpha e^{2V F_{\rm flow} t^2 \alpha^2} J_\nu (\alpha \hat{\lambda}^A) J_\nu (\alpha \hat{\lambda}^B) \\
&&\times \Big[ \frac{1}{4VF_{\rm flow}^2 t^2} \exp \left( -\frac{\hat{\lambda}^{A\, 2} + \hat{\lambda}^{B \, 2}}{8V F_{\rm flow}^2 t^2} \right) I_\nu \left( \frac{\hat{\lambda}^A\hat{\lambda}^B}{4V F_{\rm flow}^2 t^2} \right)\nn\\
&& \hspace{2cm} - \int_0^1 {\rm d} \alpha \, \alpha e^{-2V F_{\rm flow}^2 t^2 \alpha^2} J_\nu(\alpha \hat{\lambda}^A) J_\nu(\alpha \hat{\lambda}^B) \Big] \, , \nn
\ee
where $\hat{\lambda}^A=\lambda^A\Sigma V$ and $\hat{\lambda}^B=\lambda^B\Sigma V$ . 
\begin{figure}[t!]
\centerline{\includegraphics[width=0.75\textwidth]{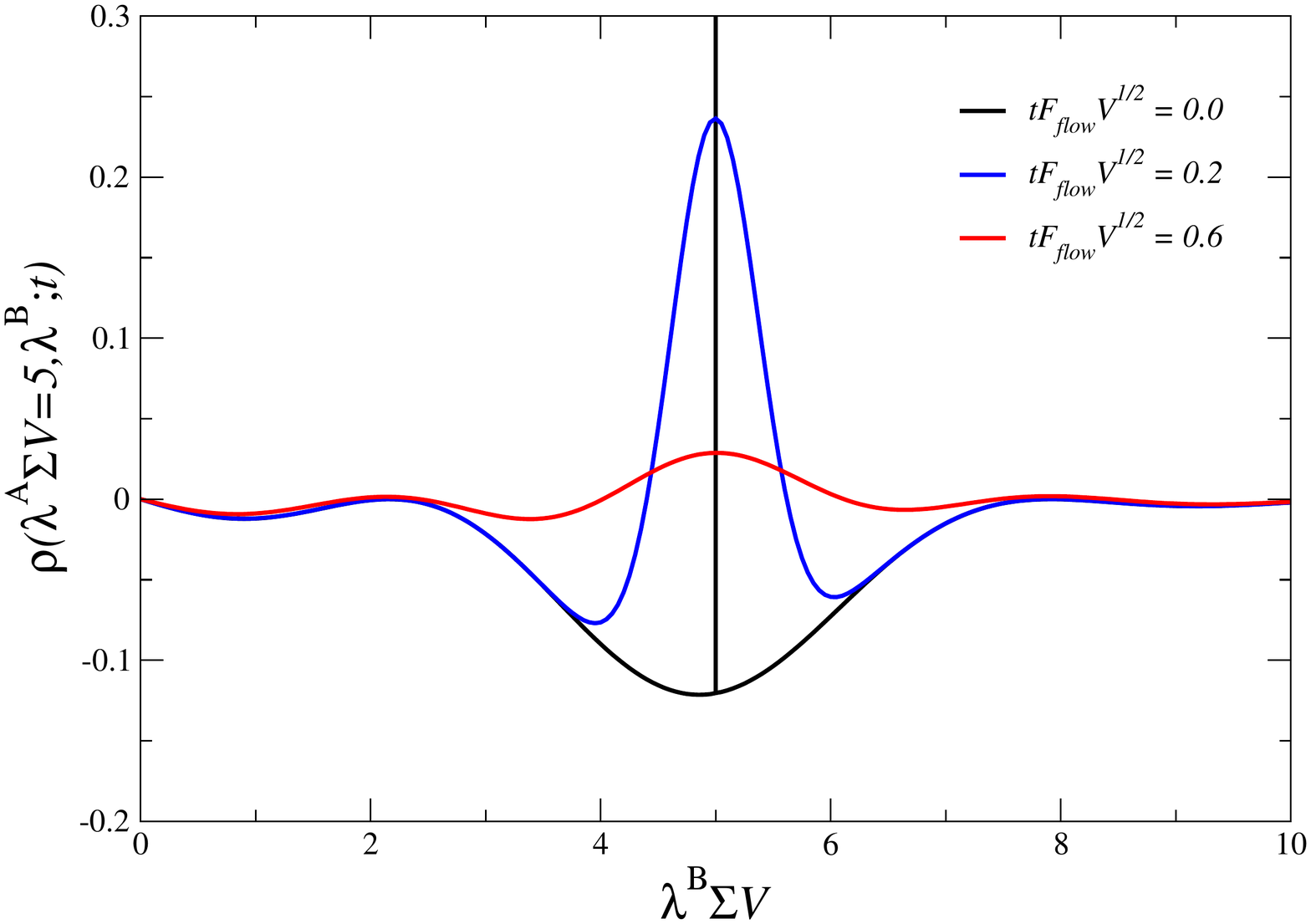}}
\caption{\label{fig:two-pfQ}The two-point correlation function between the spectrum at zero flow time and at non-zero flow time. The zero flow time eigenvalue is fixed at 
$\lambda^A\Sigma V=5$. This automatically implies that at zero flow time 
there is an eigenvalue at $\lambda^B\Sigma V=5$ and hence a $\delta$-function 
in the two-point correlation function, as indicated by the vertical line. For small flow time the primary effect of the flow is to smoothen this $\delta$-function.    
The graphs are for $\sqrt{V}F_{\rm flow}t=0.0$, $\sqrt{V}F_{\rm flow}t=0.2$ and $\sqrt{V}F_{\rm flow}t=0.6$ and $\nu=0$.}
\end{figure}
The $t\to0$ limit is also analytically accesible and agrees with the $t=0$ 
two point function \cite{TVspect}
\be
\label{twop0}
\r(\l^A,\l^B;t=0) & = & \delta(\hat{\l}^A - \hat{\l}^B) \frac{\hat{\l}^A}{2}\left[ J_\n^2(\hat{\l}^A) - J_{\n+1}(\hat{\l}^A) J_{\n-1}(\hat{\l}^A) \right]\\
&& -\frac{\hat{\l}^A\hat{\l}^B}{((\hat{\l}^A)^2 - (\hat\l^B)^2)^2} \left[ \hat\l^A J_{\n+1}(\hat{\l}^A) J_{\n}(\hat{\l}^B) - \hat{\l}^B J_{\n+1}(\hat{\l}^B) J_{\n}(\hat{\l}^A)\right]^2 \, . \nn
\ee
Note that the $\delta$-function in the first line 
is multiplied by the quenched spectral density (\ref{rhoQ}). As $t$ grows 
from zero this $\delta$-function is smeared as discussed in the previous
section. For $t$ close to zero, Eq. (\ref{twop}) can also be evaluated
analytically.
The Bessel function $I_\nu$ can be approximated by
its asymptotic form while the exponents in the integrals over $\alpha$ can be
put equal to 1. This results in
\be
\r(\l^A,\l^B;t) = \frac {\rho(\lambda_A)}{\sqrt{8\pi VF_{\rm flow}^2t^2} }  
\exp \left( -\frac{(\hat{\lambda}^{A }-  \hat{\lambda}^{B })^2}{8V F_{\rm flow}^2 t^2}\right ) +Y_2(\l^A,\l^B),
\ee
where $Y_2(\l^A,\l^B)$ represents the second term in Eq. (\ref{twop0}).
For small $t$, the width of the Gaussian distribution is given by
\be
\sigma = 2 F_{\rm flow} t \sqrt V.
\ee

The smearing of the $\delta$-function can also be seen in 
Fig.  \ref{fig:two-pfQ} where we plot
the evolution of the spectral two-point correlation function for fixed $\hat{\lambda}^A$ as a function of $\hat{\lambda}^B$ and 
$F_{\rm flow}\sqrt{V}t=0,$ 0.2 and $0.6$.
The smearing of the $\delta$-function shows that the microscopic spectrum 
of $D(B_\mu)$ decorrelates from that of $D(A_\mu)$ on the timescale set by 
$t\thicksim 1/(F_{\rm flow}\sqrt{V})$.  

The analytical form of the two point function offers an efficient 
tool to extract the value of $F_{\rm flow}$ from a given flow on the 
lattice: The width of the peak depends 
linearly 
on $F_{\rm flow}$, and a 
fit of the analytical curve to lattice data offers a way to measure 
$F_{\rm flow}$. (A similar approach has successfully been used to extract 
the value of the pion decay constant, see 
\cite{DHSS,DHSST,Damgaard:2006pu,Lehner:2009pz,AD}.)  
Different flow equations which preserve the same symmetries 
will lead to different values of $F_{\rm flow}$, and one may use the 
spectral two-point function to find the flow with the smallest value
of  $F_{\rm flow}$ and hence the flow which best preserves the dynamical 
properties of the microscopic eigenvalues.

%%%%%%%%%%%%%%%%%%%%%%%%%%%%%%%%%%%%%%%%%%%%%%%%%

\section{Conclusions and Discussion}
\label{sec:conc}

We have constructed the low energy theory 
for the gradient flow of the Dirac eigenvalues to 
leading order in the flow time. This theory 
is an extended version of partially quenched chiral 
perturbation theory, where to leading order an 
additional term proportional to $t^2$ competes with the 
mass term. The new term of order $t^2$ comes with a new 
low energy constant, $F_{\rm flow}^2$, the value of which 
depends on the detailed form of the flow equations.

Using this chiral Lagrangian we have computed the spectral 
resolvent of the Dirac operator for  gauge fields 
at non-zero flow time. 
For valence quark mass equal to the physical quark mass, it coincides
with the chiral condensate due to the flow of the gauge fields but with
no flow of the fermion fields.
The eigenvalue density of the  Dirac operator at non-zero flow time 
is given by the spectral resolvent evaluated at a purely imaginary 
valence quark mass. We have computed this eigenvalue density explicitly 
in the $\epsilon$-regime.

These results give insights in the changes with flow time of
 the dynamical properties 
of the simulation, carried out at zero flow time. Since the flow equation for the gauge 
field has no direct connection to 
the fermionic part of the action it is natural to expect that the 
characteristic features of dynamical simulations, such as eigenvalue repulsion 
from the quark mass, will be absent at  
large flow time. Indeed the results derived here for the mass dependence 
of the spectral resolvent and the unquenched eigenvalue density are 
both driven by the flow to their quenched form. The quenched results 
in the $\epsilon$-domain are obtained when the flow time satisfies 
$F^2_{\rm flow}t^2V\gg1$. The new low energy constant $F_{\rm flow}$ in 
this sense measures the degree to which the flow preserves the dynamical 
properties of the microscopic eigenvalues at small energies. Different 
flow equations will result in 
different values of $F_{\rm flow}$ which raises the question whether
 there is a systematic way to optimize the flow equations such that the 
value of $F_{\rm flow}$ is minimized and hence the dynamical properties of 
the initial configurations are better preserved. The spectral two point function, also determined here, 
offers an ideal way measure the value of $F_{\rm flow}$ in simulations. For small flow
time this constant follows immediately from the width of the peak in the two-point function.

Within the $\epsilon$-counting scheme for the spectral resolvent
we  require that  $F_{\rm flow}^2t^2V\thicksim1$ 
and only a single new term appears in the 
chiral Lagrangian at leading order. This allowed us to follow the 
evolution of the microscopic spectral observables with flow time. For a larger 
scaling of the flow time we would not have been able to resolve
this dependence.
Suppose we had allowed $t$ to be of order $1/L$, then the 
microscopic eigenvalues would have decorrelated immediately from the fermion
determinant.

A flow time of order $1/L$ is the natural scale when considering the spectral correlation 
functions in the $p$-regime of chiral perturbation theory, since the $t^2$ 
term in the chiral Lagrangian competes with the mass term for $t\thicksim 1/L$ 
(we have $m_\pi^2F_\pi^2 \thicksim t^2 F_{\rm flow}^2$ if $m_\pi \thicksim t\thicksim 1/L$). 
Using this $p$-regime counting it would be most interesting to work out the 
corrections to the Smilga-Stern relation \cite{SmilgaStern} due to the flow.

All results presented here lend themselves to a direct test in Lattice QCD. 
For an existing dynamical simulation one  needs  to compute the flow 
of the gauge fields and determine the low lying eigenvalues of the Dirac 
operator evaluated in these backgrounds. Such a computation will give
direct insights in the conservation of the dynamical properties of the 
spectral observables during the flow. Moreover it will determine if 
chiral perturbation theory captures this aspect of  gradient flow.

Finally it would be interesting to generalize the results obtained here 
for the $\epsilon$-regime of chiral perturbation theory to an arbitrary 
number of flavors.  It would also be useful to obtain explicit expressions for individual  
eigenvalue distributions. Starting from \cite{AD,AI} such a computation 
appears to be within reach.

\noindent
{\bf Acknowledgments:}
We would like to thank Poul Henrik Damgaard, Joyce C. Myers, Peter Pedersen as 
well as participants of the workshop 'Facing Strong Dynamics' for discussions. 
KS would like to thank the CERN theory division for 
hospitality during the completion of this project and the participants of the workshop 
'Conceptual advances in Lattice gauge theory' for discussions. This work was supported 
by the L{\o}rup foundation (ASC), U.S. DOE Grant No. DE-FG-88ER40388 (JV) 
and the {\sl Sapere Aude} program of The Danish Council for 
Independent Research (KS).

\appendix
\setcounter{section}{0}

%========================================================================

\section{$\langle{\cal S}^0_t\rangle$ for $N_f = 1$ by explicit parmetrization}
\label{app:parametrization}

In this appendix  we outline the derivation of $\langle{\cal S}^0_t\rangle$ for $N_f = 1$
as obtained from an explicit parameterization of the graded partition 
function at nonzero flow time
\be
Z_{2|1}(m,m_v,m_v';t) = \int_{Gl(2|1)} \hspace{-8mm}{\cal D}U \, e^{\frac{1}{2}{\rm Str}\left[{\cal  M}(U+U^{-1})\right]+t^2 F_{\rm flow}^2 V {\rm Str} \left[ {\cal T} U^{-1} {\cal T} U \right]}  \, .
\ee
Here ${\cal  M} = {\rm Diag}(m,m_v,m_v')$ and ${\cal T} = {\rm Diag}(0,1,1)$. Note that in this Appendix we have absorbed a factor of 
$\Sigma V$ into $m$, $m_v$ and $m_v'$ in order to lighten the notation.
To parameterize $U$ we choose 
\be
U &=& \left(\begin{array}{ccc} 
e^{it+iu}\cos \theta & ie^{it+i\phi}\sin\theta & 0 \\ 
ie^{it-i\phi}\sin\theta & e^{it-iu}\cos\theta & 0 \\ 
0 & 0 & e^{s} \end{array}\right)
\exp\left(\begin{array}{ccc}
0 & 0 & \al_1 \\ 
0 & 0 & \al_2 \\ 
\beta_1 & \beta_2 & 0 \end{array}\right) \, ,
\label{U}
\ee
where $\theta,t,u\in[-\pi,\pi]$, $\phi\in[0,\pi]$ and $s\in[-\infty,\infty]$. The corresponding Jacobian and Berezinian are \cite{DOTV} 
\be
J = 4 e^{4it}\cos \theta\sin \theta \, , \qquad B = 1+\frac{1}{3}(\alpha_1\beta_1+\alpha_2\beta_2) \, .
\ee
The Grassmann integrals can be carried out analytically and result in the partition function
\be
&& Z_{2|1}(m,m_v,m_v';t)\\
&=&\int_{-\infty}^\infty {\rm d}s \int_0^\pi {\rm d}\phi \int_{-\pi}^\pi {\rm d}t
\int_{-\pi}^\pi {\rm d}u \int_{-\pi}^\pi {\rm d}\theta \ |J| \  (P_4^t-P_{11}^tP_{22}^t+P_{12}^tP_{21}^t-\frac{1}{3}P_{11}^t-\frac{1}{3}P_{22}^t) \ e^{S^t} , \nn
\ee
where
\begin{equation}
S^t = m \cos(t+u) \cos\theta + m_v \cos(t-u) \cos\theta - m_v'\cosh s - t^2 F_{\rm flow}^2 V \sin^2\theta \, ,
\end{equation}
and
\be
P_{11}^t &=&\, \frac{m}{2} \cos(t+u) \cos\theta + \frac{m_v'}{2} \cosh s + t^2 F_{\rm flow}^2 V \cos^2\theta \, , \nn \\
P_{22}^t &=&\, \frac{m_v}{2} \cos(t-u) \cos\theta + \frac{m_v'}{2} \cosh s \, , \nn \\
P_{12}^t &=&\, -\frac{i m}{4} e^{-it-i\phi} \sin\theta + \frac{i m_v}{4} e^{it-i\phi}\sin\theta - t^2 F_{\rm flow}^2 V \tfrac{i}{2}\cos\theta \sin\theta e^{iu-i\phi} \, , \label{Pterms} \\
P_{21}^t &=&\, \frac{i m}{4} e^{it + i\phi}\sin\theta - \frac{i m_v}{4}e^{-it+i\phi}\sin\theta + t^2 F_{\rm flow}^2 V \tfrac{i}{2}\cos\theta \sin\theta e^{-iu+i\phi} \, , \nn \\
P_{4}^t &=&\, \frac{m}{24} \cos(t+u) \cos\theta + \frac{m_v}{24} \cos(t-u) \cos\theta + \frac{m_v'}{12} \cosh s+ t^2 F_{\rm flow}^2 V \frac{1 + 2 \sin^2 \theta}{12}  \, . \nn
\ee
The $\phi$ dependence of the integrand cancels, and thus the $\phi$-integration trivially yields an overall factor of $\pi$. It is furthermore possible to express the $s$-integration in terms of modified Bessel functions of the second kind, using
\be
 \int_{-\infty}^\infty \hspace{-3mm}{\rm d}s \, e^{-m \cosh s} &=& 2 K_0(m) \, , \nn \\
 \int_{-\infty}^\infty \hspace{-3mm}{\rm d}s  \, e^{-m \cosh s}\cosh s &=& 2 K_1(m) \, , \\
 \int_{-\infty}^\infty \hspace{-3mm}{\rm d}s \,  e^{-m \cosh s} \cosh^2 s &=& K_0(m) + K_2(m)  \, . \nn
\ee
This allows for writing the generation functional as
\be
&& Z_{2|1}(m,m_v,m_v';t) = \pi \int_{-\pi}^\pi {\rm d}t \int_{-\pi}^\pi {\rm d}u \int_{-\pi}^\pi {\rm d}\theta 
\, |J| \,  (P_0^t + P_1^t + P_2^t) \, e^{S_{\rm int}^t}  \, ,
\ee 
where
\be
S_{\rm int}^t = m \cos(t+u) \cos\theta + m_v \cos(t-u) \cos\theta - t^2 F_{\rm flow}^2 V \sin^2\theta  \, ,
\ee 
and
\be
P_0^t &=&\, 2 K_0(m_v') \Bigg[ P_{12}^t P_{21}^t \nn\\
&&+ \bigg\{ \frac{m}{24} \cos(t+u) \cos\theta + \frac{m_v}{24} \cos(t-u) \cos\theta - t^2 F_{\rm flow}^2 V \frac{1 + 2 \sin^2 \theta}{12}\bigg\}_4\nn\\
&&- \bigg\{ \Big(\frac{m}{2} \cos(t+u) \cos\theta  + t^2 F_{\rm flow}^2 V \cos^2\theta \Big)\Big( \frac{m_v}{2} \cos(t-u) \cos\theta \Big) \bigg\}_{\substack{11\\22}} \\
&&-\frac{1}{3}  \bigg\{ \frac{m}{2} \cos(t+u) \cos\theta + t^2 F_{\rm flow}^2 V \cos^2\theta \bigg\}_{11} -\frac{1}{3}  \bigg\{  \frac{m_v}{2} \cos(t-u) \cos\theta  \bigg\}_{22}  \Bigg] \, ,\nn\\
P_1^t &=&\, -\frac{m_v'}{2} K_1(m_v') \left[ 1+m_v \cos(t-u) \cos\theta - m \cos(t+u) \cos\theta - 2 t^2F_{\rm flow}^2 V \cos^2\theta \right] \, ,\nn\\
P_2^t &=&\, -\frac{m_v'}{4} [K_0(m_v') + K_2(m_v')] \, . \nn
\ee
The subscripts on the curly brackets are only reminders of where the term originates from in \eqref{Pterms}.

The parameterization allows for a semi-analytical calculation of the spectral resolvent as a function of $m$ and at non-zero flow time from
\be
\langle{\cal S}^0_t\rangle_{m, m_v;t} = \frac{1}{Z_{2|1}(m,m_v,m_v;t)} \, \partial_{m_v} Z_{2|1}(m,m_v,m_v';t) \bigg\vert_{m_v = m_v'} \, .
\ee
Note that $Z_{2|1}(m,m_v,m_v;t)=Z_{1}(m)$.

%==================================================================


\begin{thebibliography}{99}
\bibitem {luscher-1}
%\cite{Luscher:2009eq}
%\bibitem{Luscher:2009eq} 
  M.~L\"uscher,
  %``Trivializing maps, the Wilson flow and the HMC algorithm,''
  Commun.\ Math.\ Phys.\  {\bf 293}, 899 (2010)
  [arXiv:0907.5491 [hep-lat]].
  %%CITATION = ARXIV:0907.5491;%%
  %38 citations counted in INSPIRE as of 27 May 2014


%\cite{Luscher:2010iy}
\bibitem{luscher-2}
%\bibitem{Luscher:2010iy} 
  M.~L\"uscher,
  %``Properties and uses of the Wilson flow in lattice QCD,''
  JHEP {\bf 1008}, 071 (2010)
  [arXiv:1006.4518 [hep-lat]].
  %%CITATION = ARXIV:1006.4518;%%
  %83 citations counted in INSPIRE as of 27 May 2014

%\cite{Luscher:2011bx}
\bibitem{luscher-weisz}
%\bibitem{Luscher:2011bx} 
  M.~L\"uscher and P.~Weisz,
  %``Perturbative analysis of the gradient flow in non-abelian gauge theories,''
  JHEP {\bf 1102}, 051 (2011)
  [arXiv:1101.0963 [hep-th]].
  %%CITATION = ARXIV:1101.0963;%%
  %33 citations counted in INSPIRE as of 27 May 2014

\bibitem{luscher-chiral}
%\cite{Luscher:2013cpa}
%\bibitem{Luscher:2013cpa} 
  M.~L\"uscher,
  %``Chiral symmetry and the Yang--Mills gradient flow,''
  JHEP {\bf 1304}, 123 (2013)
  [arXiv:1302.5246 [hep-lat]].
  %%CITATION = ARXIV:1302.5246;%%
  %17 citations counted in INSPIRE as of 27 May 2014

\bibitem{Suzuki} 
  H.~Makino and H.~Suzuki,
  %``Lattice energy?momentum tensor from the Yang?Mills gradient flow?inclusion of fermion fields,''
  PTEP {\bf 2014}, no. 6, 063B02 (2014)
  [arXiv:1403.4772 [hep-lat]].
  %%CITATION = ARXIV:1403.4772;%%

\bibitem{DelDPR} 
  L.~Del Debbio, A.~Patella and A.~Rago,
  %``Space-time symmetries and the Yang-Mills gradient flow,''
  JHEP {\bf 1311}, 212 (2013)
  [arXiv:1306.1173 [hep-th]].
  %%CITATION = ARXIV:1306.1173;%%

\bibitem{Daniel} 
  Z.~Fodor, K.~Holland, J.~Kuti, S.~Mondal, D.~Nogradi and C.~H.~Wong,
  %``The lattice gradient flow at tree-level and its improvement,''
  arXiv:1406.0827 [hep-lat].
  
  \bibitem{RS}
  A.~Ramos and S.~Sint, {\sl talk at 'Conceptual advances in lattice gauge theory (LGT14)', CERN July 2014.} 
  
  
\bibitem{DN} 
  P.~H.~Damgaard and S.~M.~Nishigaki,
  %``Universal spectral correlators and massive Dirac operators,''
  Nucl.\ Phys.\ B {\bf 518}, 495 (1998)
  [hep-th/9711023].
  %%CITATION = HEP-TH/9711023;%%
  %93 citations counted in INSPIRE as of 22 Jul 2014

\bibitem{WGW} 
  T.~Wilke, T.~Guhr and T.~Wettig,
  %``The Microscopic spectrum of the QCD Dirac operator with finite quark masses,''
  Phys.\ Rev.\ D {\bf 57}, 6486 (1998)
  [hep-th/9711057].
  %%CITATION = HEP-TH/9711057;%%
  
\bibitem{DOTV} 
  P.~H.~Damgaard, J.~C.~Osborn, D.~Toublan and J.~J.~M.~Verbaarschot,
  %``The microscopic spectral density of the QCD Dirac operator,''
  Nucl.\ Phys.\ B {\bf 547}, 305 (1999)
  [hep-th/9811212].
  %%CITATION = HEP-TH/9811212;%%


\bibitem{BerbenniBitsch:1998sy}
  M.~E.~Berbenni-Bitsch, S.~Meyer and T.~Wettig,
  %``Microscopic universality with dynamical fermions,''
  Phys.\ Rev.\ D {\bf 58}, 071502 (1998)
  [hep-lat/9804030].
  %%CITATION = HEP-LAT/9804030;%%

\bibitem{Damgaard:2000qt} 
  P.~H.~Damgaard, U.~M.~Heller, R.~Niclasen and K.~Rummukainen,
  %``Eigenvalue distributions of the QCD Dirac operator,''
  Phys.\ Lett.\ B {\bf 495}, 263 (2000)
  [hep-lat/0007041].
  %%CITATION = HEP-LAT/0007041;%%

\bibitem{Fukaya} 
  H.~Fukaya {\it et al.}  [JLQCD and TWQCD Collaborations],
  %``Determination of the chiral condensate from QCD Dirac spectrum on the lattice,''
  Phys.\ Rev.\ D {\bf 83}, 074501 (2011)
  [arXiv:1012.4052 [hep-lat]].
  %%CITATION = ARXIV:1012.4052;%%

\bibitem{BG}
%\cite{Bar:2013ora}
%\bibitem{Bar:2013ora} 
  O.~Bar and M.~Golterman,
  %``Chiral perturbation theory for gradient flow observables,''
  Phys.\ Rev.\ D {\bf 89}, 034505 (2014)
  [arXiv:1312.4999 [hep-lat]].
  %%CITATION = ARXIV:1312.4999;%%
  %2 citations counted in INSPIRE as of 27 May 2014

\bibitem{shindler}
%\cite{Shindler:2013bia}
%\bibitem{Shindler:2013bia} 
  A.~Shindler,
  %``Chiral Ward identities, automatic O(a) improvement and the gradient flow,''
  Nucl.\ Phys.\ B {\bf 881}, 71 (2014)
  [arXiv:1312.4908 [hep-lat]].
  %%CITATION = ARXIV:1312.4908;%%
  %1 citations counted in INSPIRE as of 27 May 2014

%\bibitem{Luscher1} 
%  M.~Luscher,
  %``Properties and uses of the Wilson flow in lattice QCD,''
%  JHEP {\bf 1008}, 071 (2010)
%  [arXiv:1006.4518 [hep-lat]].
  %%CITATION = ARXIV:1006.4518;%%

%\bibitem{Luscher2} 
%  M.~Luscher,
  %``Chiral symmetry and the Yang--Mills gradient flow,''
%  JHEP {\bf 1304}, 123 (2013)
%  [arXiv:1302.5246 [hep-lat]].
  %%CITATION = ARXIV:1302.5246;%%

\bibitem{GLorig} 
  J.~Gasser and H.~Leutwyler,
  %``Chiral Perturbation Theory to One Loop,''
  Annals Phys.\  {\bf 158}, 142 (1984).
  %%CITATION = APNYA,158,142;%%

\bibitem{GL}
  J.~Gasser and H.~Leutwyler,
  %``Thermodynamics of Chiral Symmetry,''
  Phys.\ Lett.\ B {\bf 188}, 477 (1987).
  %%CITATION = PHLTA,B188,477;%%LS

\bibitem{LS}
  H.~Leutwyler and A.~V.~Smilga,
  %``Spectrum of Dirac operator and role of winding number in QCD,''
  Phys.\ Rev.\ D {\bf 46}, 5607 (1992).
  %%CITATION = PHRVA,D46,5607;%%


\bibitem{DHSST} 
  P.~H.~Damgaard, U.~M.~Heller, K.~Splittorff, B.~Svetitsky and D.~Toublan,
  %``Microscopic eigenvalue correlations in QCD with imaginary isospin chemical potential,''
  Phys.\ Rev.\ D {\bf 73}, 105016 (2006)
  [hep-th/0604054].
  %%CITATION = HEP-TH/0604054;%%

\bibitem{VZ} 
  J.~J.~M.~Verbaarschot and I.~Zahed,
  %``Spectral density of the QCD Dirac operator near zero virtuality,''
  Phys.\ Rev.\ Lett.\  {\bf 70}, 3852 (1993)
  [hep-th/9303012].
  %%CITATION = HEP-TH/9303012;%%

\bibitem{DHSS} 
  P.~H.~Damgaard, U.~M.~Heller, K.~Splittorff and B.~Svetitsky,
  %``A New method for determining F(pi) on the lattice,''
  Phys.\ Rev.\ D {\bf 72}, 091501 (2005)
  [hep-lat/0508029].
  %%CITATION = HEP-LAT/0508029;%%


\bibitem{Damgaard:2006pu} 
  P.~H.~Damgaard, U.~M.~Heller, K.~Splittorff, B.~Svetitsky and D.~Toublan,
  %``Extracting F(pi) from small lattices: Unquenched results,''
  Phys.\ Rev.\ D {\bf 73}, 074023 (2006)
  [hep-lat/0602030].
  %%CITATION = HEP-LAT/0602030;%%


\bibitem{SVtoda} 
  K.~Splittorff and J.~J.~M.~Verbaarschot,
  %``Replica limit of the Toda lattice equation,''
  Phys.\ Rev.\ Lett.\  {\bf 90}, 041601 (2003)
  [cond-mat/0209594].
  %%CITATION = COND-MAT/0209594;%%


\bibitem{rep-sk}
D. Sherrington and S. Kirkpatrick, 
%Solvable Model of a Spin-Glass, 
Phys. Rev. Lett. {\bf 35}, 1792 (1975).


\bibitem{replica} 
  P.~H.~Damgaard and K.~Splittorff,
  %``Partially quenched chiral perturbation theory and the replica method,''
  Phys.\ Rev.\ D {\bf 62}, 054509 (2000)
  [hep-lat/0003017].
  %%CITATION = HEP-LAT/0003017;%%



\bibitem{Kanzieper} 
  E.~Kanzieper,
  %``Random matrices and the replica method,''
  Nucl.\ Phys.\ B {\bf 596}, 548 (2001)
  [cond-mat/9908130].
  %%CITATION = COND-MAT/9908130;%%


\bibitem{SVfact} 
  K.~Splittorff and J.~J.~M.~Verbaarschot,
  %``Factorization of correlation functions and the replica limit of the Toda lattice equation,''
  Nucl.\ Phys.\ B {\bf 683}, 467 (2004)
  [hep-th/0310271].
  %%CITATION = HEP-TH/0310271;%%


\bibitem{TVspect} 
  D.~Toublan and J.~J.~M.~Verbaarschot,
  %``Statistical properties of the spectrum of the QCD Dirac operator at low-energy,''
  Nucl.\ Phys.\ B {\bf 603}, 343 (2001)
  [hep-th/0012144].
  %%CITATION = HEP-TH/0012144;%%

\bibitem{Lehner:2009pz} 
  C.~Lehner and T.~Wettig,
  %``Partially quenched chiral perturbation theory in the epsilon regime at next-to-leading order,''
  JHEP {\bf 0911}, 005 (2009)
  [arXiv:0909.1489 [hep-lat]].
  %%CITATION = ARXIV:0909.1489;%%

\bibitem{AD} 
  G.~Akemann and P.~H.~Damgaard,
  %``Individual Eigenvalue Distributions of Chiral Random Two-Matrix Theory and the Determination of F(pi),''
  JHEP {\bf 0803}, 073 (2008)
  [arXiv:0803.1171 [hep-th]].
  %%CITATION = ARXIV:0803.1171;%%


\bibitem{SmilgaStern} 
A.~V.~Smilga and J.~Stern,
  %``On the spectral density of Euclidean Dirac operator in QCD,''
  Phys.\ Lett.\ B {\bf 318}, 531 (1993).
  %%CITATION = PHLTA,B318,531;%%


\bibitem{AI} 
  G.~Akemann and A.~C.~Ipsen,
  %``The k-th Smallest Dirac Operator Eigenvalue and the Pion Decay Constant,''
  J.\ Phys.\ A {\bf 45}, 115205 (2012)
  [arXiv:1110.6774 [hep-lat]].
  %%CITATION = ARXIV:1110.6774;%%

\end{thebibliography}
\end{document}